# Promoting the Joy of Learning by Turning a Smartphone into Scientific Equipment

*Tze Kwang LEONG[1], Loo Kang WEE[1], Félix J. GARCÍA_CLEMENTE[2] & Francisco ESQUEMBRE[3]*

[1]*Ministry of Education, Singapore*

[2]*Department of Computer Engineering and Technology, University of Murcia, Spain*

[3]*Department of Mathematics, University of Murcia, Spain*

**Abstract**. Smartphone is a powerful internet connected computer packed with internal sensors that measure sound, light, acceleration and magnetic field strength. Physics teachers can use them as measurement devices to demonstrate science concepts and promote the joy of learning. Some research has also shown the benefits of using smartphones for teaching and learning physics. This article aims to extend mobile phone research with our open source apps and low-cost experimental design. The three experiments that we designed include radioactivity, line spectrum, standing sound wave and polarization of light. Students are able to conduct experiments and collect data easily without using bulky data loggers or laptops. Substituting a data logger with a smartphone will mean that every student can possess his/her own measuring tool inside and outside the classroom, instead of having to share with a large group of students in a scheduled lab. By empowering the students to conduct their own experiments and collect data, the use of the smartphone as a tool aims to support student learning anytime and anywhere, igniting greater joy of learning.

## 1 Introduction

Physics is an experimental science based on valid experiential evidence (Dewey, 1958), criticism, and rational discussion. Smartphones (Briggle, 2013; Hall, 2013; Monteiro, P., Stari, C., & Martí, 2016; Monteiro et al., 2017; Sans, Manjón, Pereira, Gomez-Tejedor, & Monsoriu, 2013) come packed with a variety of sensors such as accelerometer, gyroscope, light, microphone, camera and touch screen, making them ideal as an experiment tool. There are many free Physics mobile applications such as Physics Toolbox Sensor Suite by Vieyra Software and phyphox by RWTH Aachen University that can collect and record data using the phone sensors. Many current research (Kuhn & Vogt, 2013) focuses on using the sensors directly such as the accelerometer and gyroscope motion sensor. However, experiments for waves, quantum physics and nuclear physics have not received as much attention.

Thus, in this paper, we discuss three experimental applications' design to study standing waves, line spectrum and radioactivity.

The experiments were used for teaching students between Grades 11 and 12.

## 2 Design of Smartphone experiments

Each of the three experiments comes with a custom design web app, Android and iOS app so that the students can easily conduct the experiment and collect real-time experimental evidence. These sensor aware apps (Esquembre, García Clemente, & Wee, 2018) are created using Easy JavaScript Simulation (Wolfgang Christian & Esquembre, 2012; W. Christian, Esquembre, & Mason, 2010; Esquembre, 2004, 2012; Wee, 2010, 2012, 2016) (EJSS), an multi award winning authoring toolkit[1]. We recommend in order of compatibility, web apps using just the Google Chrome internet browser, iOS Safari and install the Android app. The iOS app is currently less developed, but we hope to publish a fully featured iOS app as soon as possible when the technology is supported on Xcode.

a) **Sound analyzer app[2]:**

By blowing into PVC pipes of different length and diameter as shown in *Figure 1*, students can hear the difference in pitch of the sound produced by the pipe. The pitch also changes when the pipe is closed. However, it is tedious for them to study the sound quantitatively.

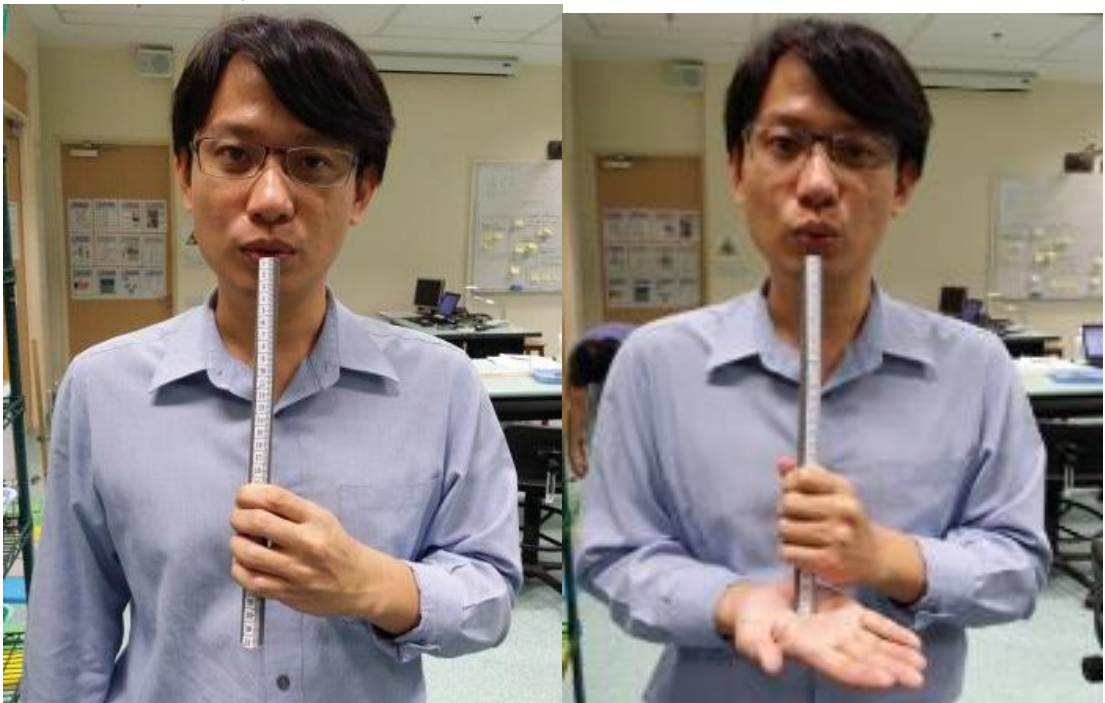

**Figure 1.** *Blowing into the PVC pipe with both ends open (left) and closed on one end of the pipe with the palm (right)*

Students are able to use our app to observe the resonant frequencies of the pipe and model the relationship of standing waves of both end open pipe and one end closed pipe. Our app analyzes the sound from the smartphone microphone using Fourier transform and presents the data in the form of intensity versus frequency graph as shown in *Figure 2 and 3*. The app also predicts the length of the pipe as well as determine whether the pipe is open at both ends or closed at one end.

---

[1] https://www.um.es/fem/EjsWiki/

[2] gg.gg/soundapp

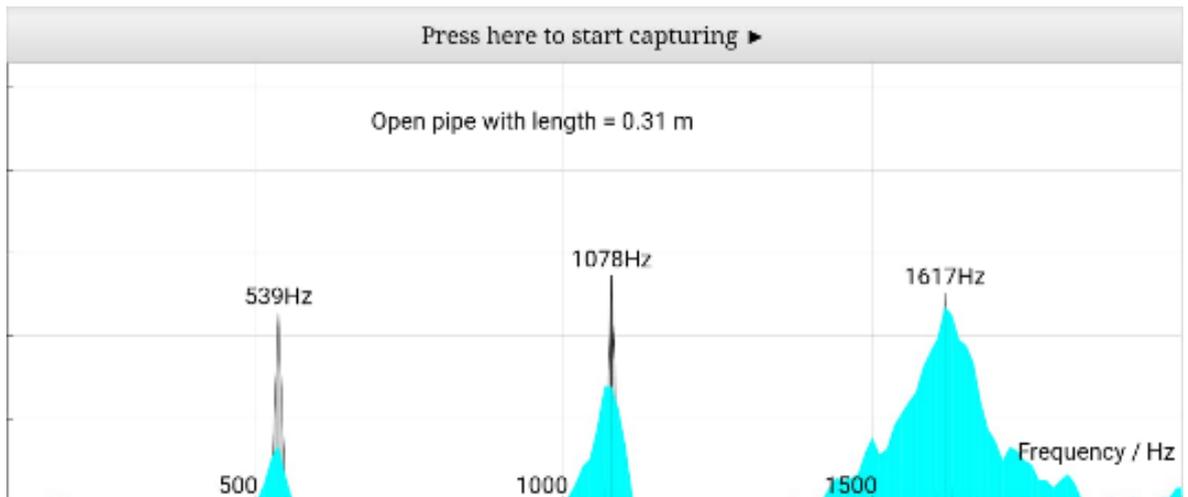

*Figure 2.* Screenshot of the web app correctly predicting that the pipe is open at both ends and that the length of the pipe is 0.31 m.

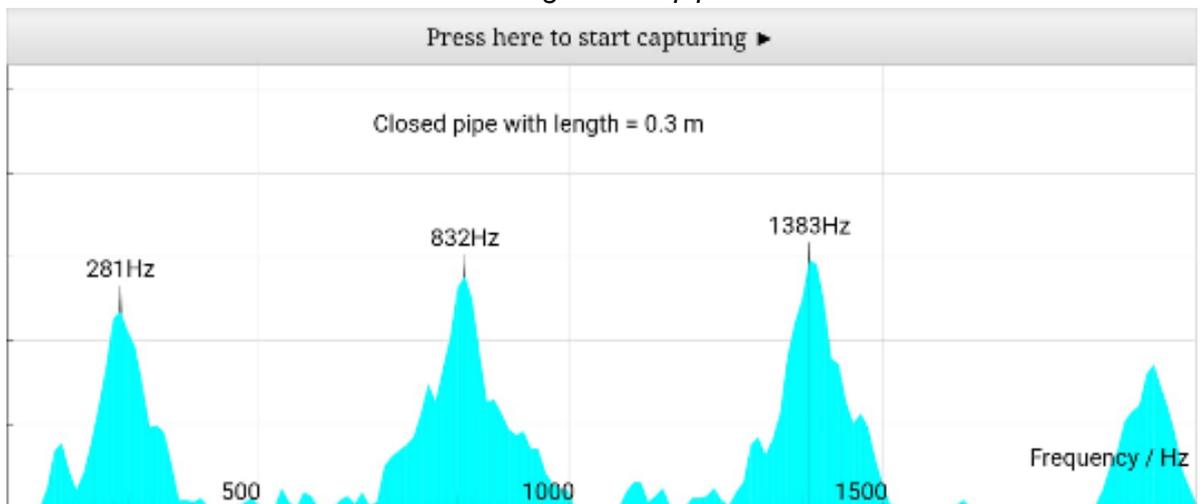

*Figure 3.* Screenshot of the app correctly predicting that the pipe is closed at one end and that the length of the pipe is 0.31 m.

Students will record the harmonic resonant frequencies of the pipes and determine the corresponding wavelength, λ using the formula

$$v = f \lambda$$

and assuming velocity of sound in air, *v* is 330 m/s as shown in *Table 1* and *Table 2*. Students will then deduce the relationship between the wavelength of the pipe and the length of the pipe.

| Order of harmonics | Open Pipe frequency /Hz | Open Pipe wavelength, λ/ m | Length of pipe/Wavelength, L/λ |
|---|---|---|---|
| 1 | 539 | 0.61 | 1/2 |
| 2 | 1078 | 0.31 | 1 |
| 3 | 1617 | 0.20 | 3/2 |

*Table 1. Data of Open Pipe showing the relationship between frequency, wavelength and the length of the pipe (L =0.30 m long)*

Students will be able to observe that the pattern for pipe closed in one end is different from the pipe that is open at both ends as shown in *Table 2*.

| Order of harmonics | Closed Pipe frequency /Hz | Closed Pipe wavelength, λ/ m | Length of pipe/Wavelength, L/λ |
|---|---|---|---|
| 1 | 281 | 1.17 | 1/4 |
| 2 | 832 | 0.40 | 3/4 |
| 3 | 1383 | 0.24 | 5/4 |

*Table 2. Data of Pipe Closed at one end showing the relationship between frequency, wavelength and the length of the pipe (L =0.30 m long)*

We created a (Aguirregabiria & Wee, 2011) simulation[3] (as shown in *Figure 4 to 6*) to help students understand the relationship between wavelength and the length of the pipe as well as the effect of the pipe being both end open and one end closed. Students will observe that the particles are at the open ends are displacement antinodes while particles at the closed ends are displacement nodes. From the displacement-position graph, they can see how the wavelength is related to the length of the pipe.

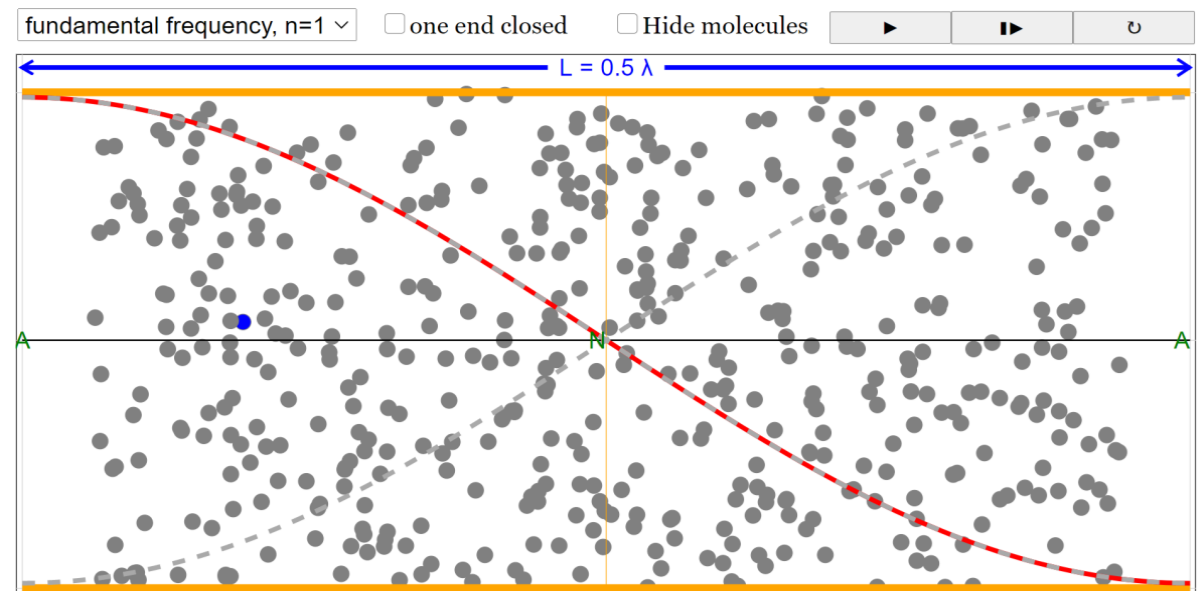

**Figure 4**. *Screenshot of the simulation to show the particle displacement for n = 1 for pipe open at both ends. Red line indicates the displacement-position graph of the particle*

[3] gg.gg/wavepipe

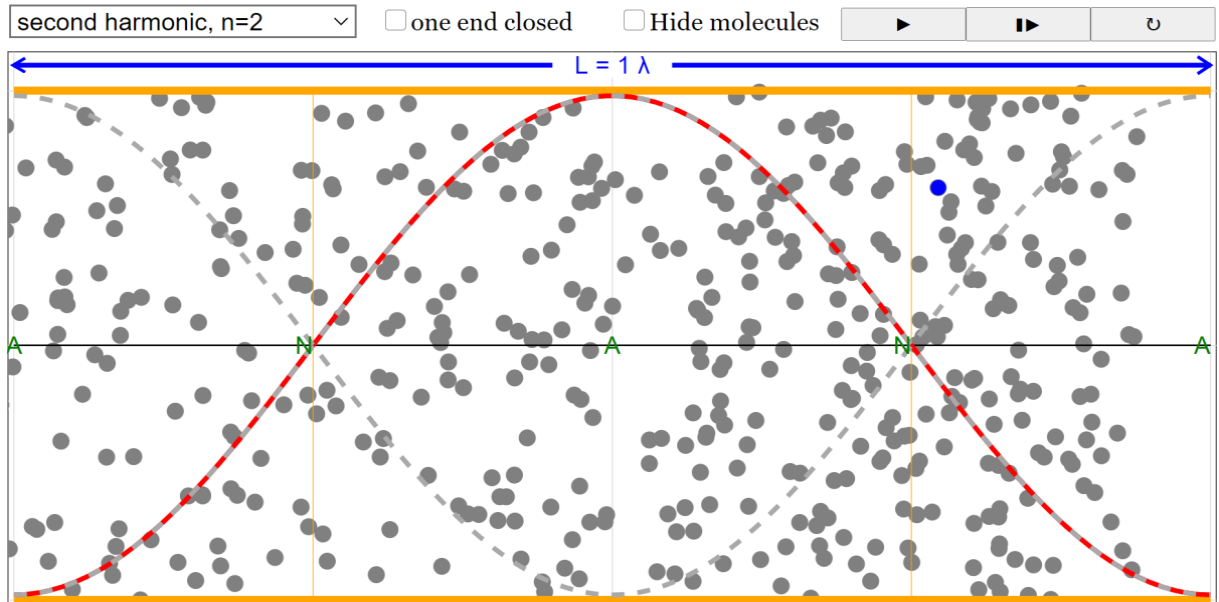

*Figure 5. Screenshot of the simulation to show the particle displacement for n = 2 for pipe open at both ends*

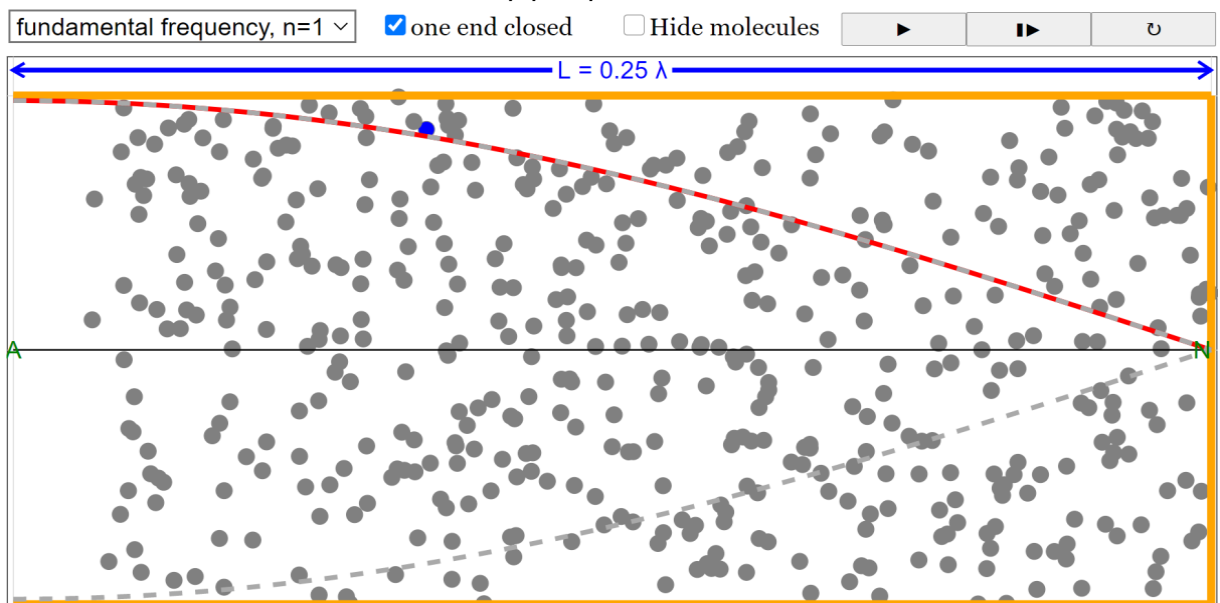

*Figure 6. Screenshot of the simulation to show the particle displacement for n = 1 for pipe closed at one end.*

**b) Line spectrum [4]app:**

Students can observe spectral lines by viewing a light produced by a gas lamp through a diffraction grating and even use video analysis (Brown & Cox, 2009). However, it could be time consuming for them to study the lines quantitatively.

Using a 3-D printed spectrometer (details of the spectrometer can be found in appendix A) as shown *in Figure 8* and our app, students can easily analyze the light and display it as intensity against position graph as shown in *Figure 9*.

---

[4] gg.gg/lightapp

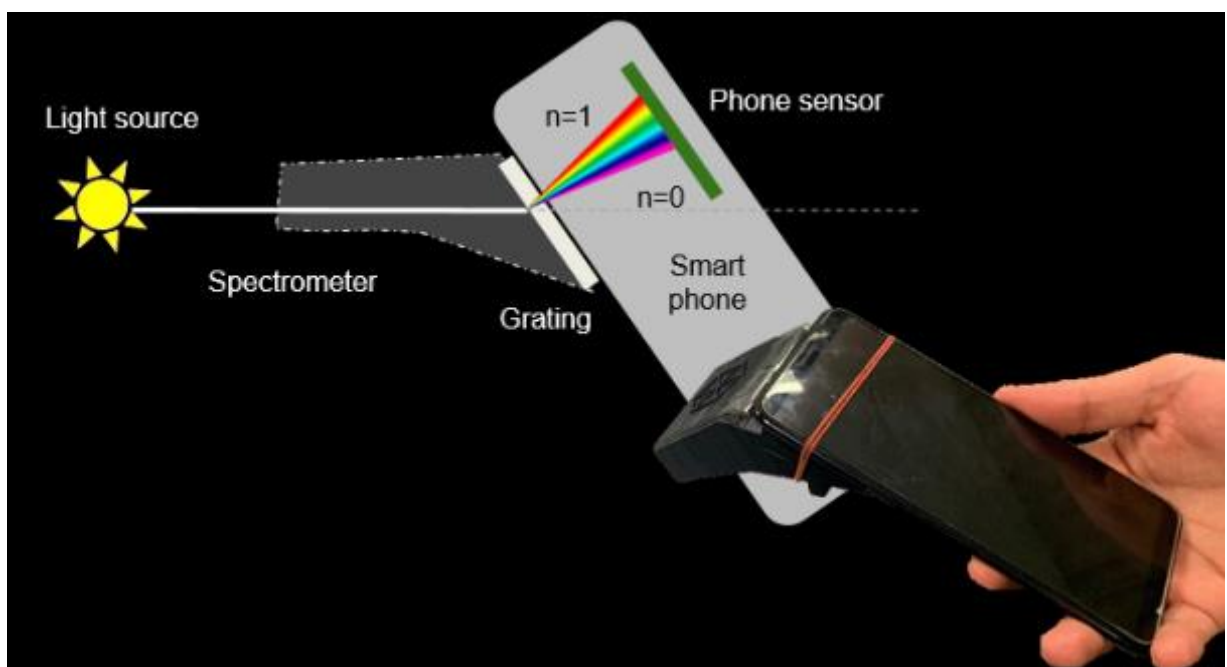

*Figure 8. The photo shows how the spectrometer is attached to a smartphone. The ray diagram illustrates how the first order diffraction pattern is being captured by the phone camera.*

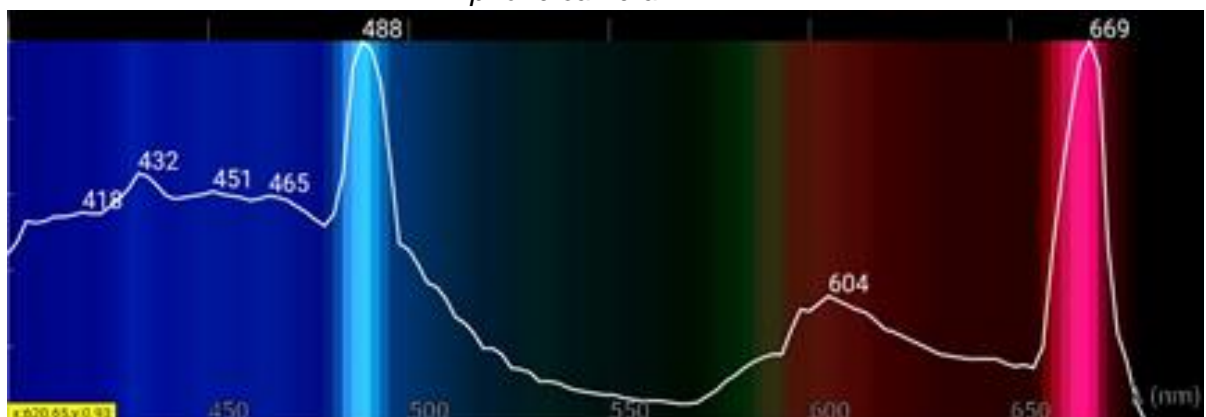

*Figure 9. Screenshot of the app measuring the line spectrum from a hydrogen lamp*

As every camera is different, students need to calibrate their camera to get an accurate measurement of the spectrum. Currently we use fluorescent light as calibration because it is easily found in classrooms in Singapore. The app will identify the red and green lines of the fluorescent light and assign them with the values 611.6 and 546.5 nm obtained from National Institute of Standards and Technology (NIST) USA. The spectrometer is calibrated by pointing the spectrometer at a blank white paper that is reflecting light from the ceiling fluorescent lamp as shown in *Figure 10*.

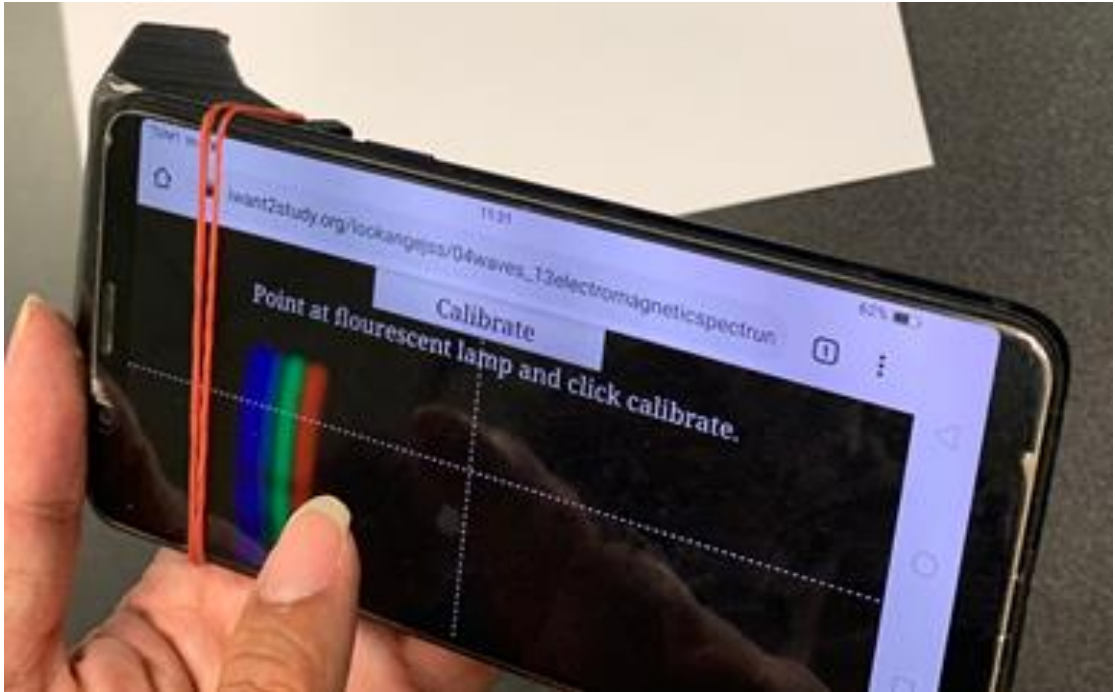

***Figure 10***. *Calibrating the spectrometer by pointing the camera at a blank white paper that is reflecting light from the ceiling fluorescent lamp*

Once calibrated, the app can be used to measure the wavelength of various light sources as shown in *Figure 9*.

We have also made the app able to identify certain gases (Helium, Neon, Mercury, Hydrogen) as shown in *Figure 11*. The app only contains the database of these 4 gases and any other gas may be wrongly identified as one of the 4 gases. This version can be accessed at gg.gg/lightapppro

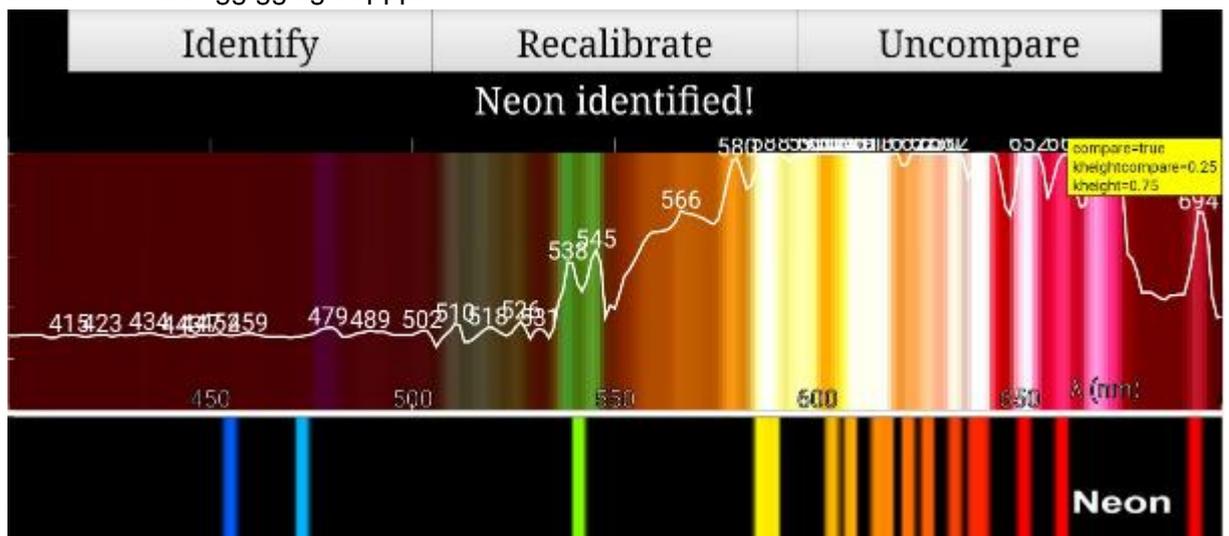

***Figure 11***. *Screenshot of the app identifying the Neon gas line spectrum (top) and comparing it to the NIST data (bottom)*

c) Radioactivity [5]app:

The third app is meant for students to measure radioactivity of beta and gamma sources using a pin-diode as a GM counter.

By inserting a pin-diode[6] into the microphone/headphone jack, the app is then able to detect beta and gamma particles as electrical signals and process them into count per minute graph and count per second against time graph as shown in *Figure 12*.

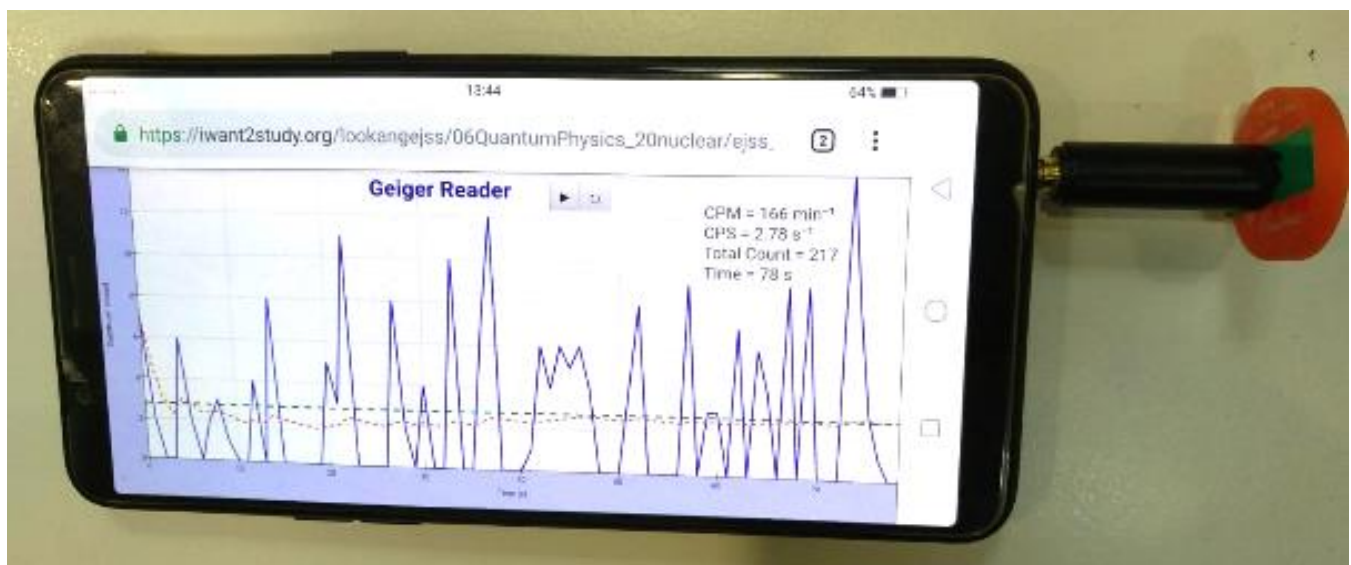

***Figure 12***. *Count per second graph blue lines of detected beta particles detected from Cobalt-60 source*

Students are able to better appreciate the random nature of the radiation by observing the instantaneous beta and gamma radiation track by the blue line collected as shown in *Figure 12*.

By placing different materials such as paper, aluminum foil between the source and the pin-diode, students can compare the penetration power of different radiation.

Students can also study how the activity changes with distance from the source.

Because of the low cost of the pin-diode, students can place multiple phones, each with a pin-diode detector, at different positions to see how the beta particles from the Cobalt-60 sources are being deflected by a neodymium magnet as shown in *Figure 13*.

---

[5] gg.gg/gmapp
[6] gg.gg/pindiodepurchase

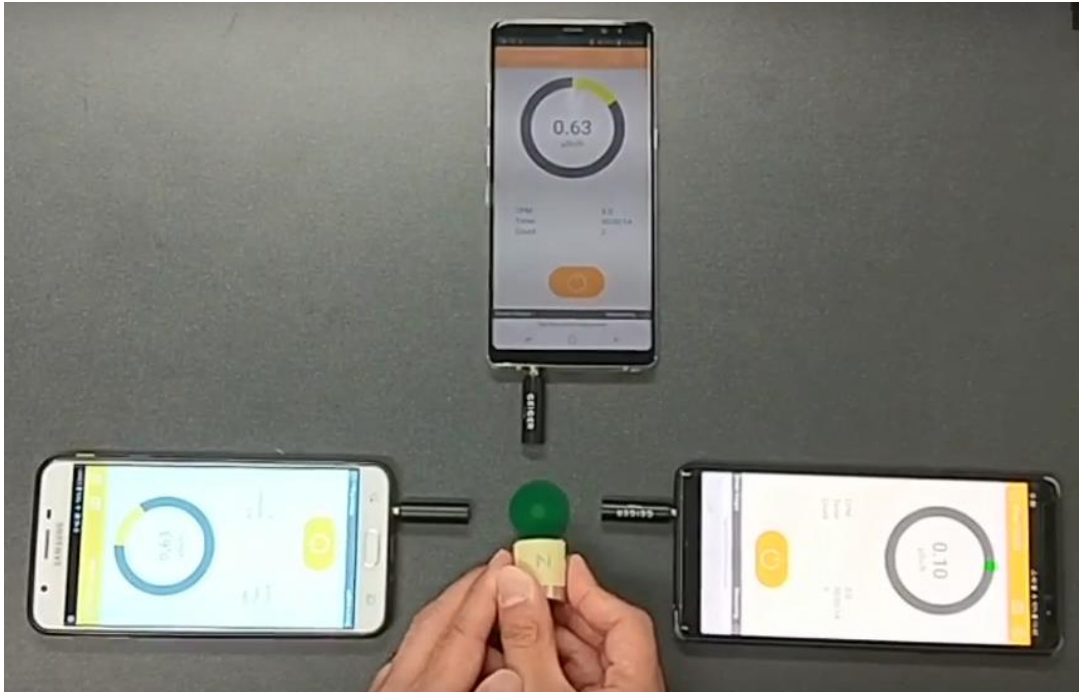

*Figure 13. The neodymium magnet deflects the beta particles to the left causing the radiation detected by the phone on the right to be less than the radiation detected by the phone on the left.*

Another experiment that can be conducted is to differentiate contamination with irradiation. Students were asked to place the radiation source on an object for a duration and predict whether there will be any change in the radioactivity on the object. Most of the students were surprised to know that there is no change in the radioactivity of the object.

Radioactive sources may not be commonly accessible to schools. Low activity Amercium-241 can be extracted from smoke detector as shown in *Figure 14*. The diode-pin is able to detect the gamma particle but not the alpha particle. It is still sufficient to describe the experiments stated above.

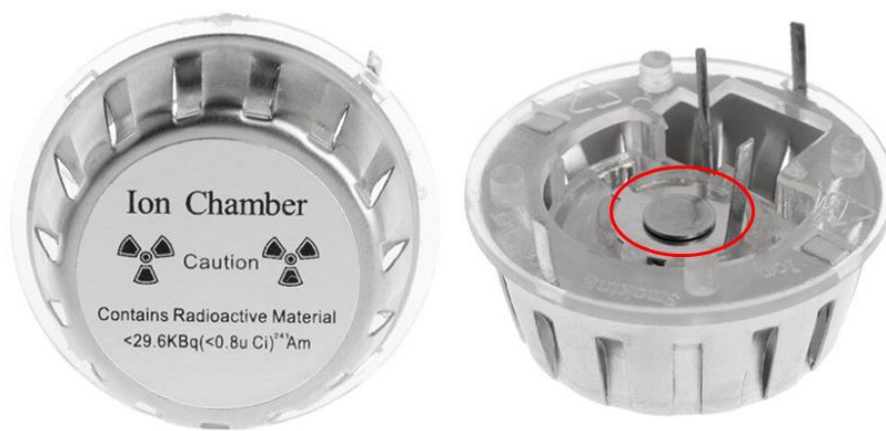

*Figure 14. Ion Chamber of Smoke detector can be easily purchased online. Low activity Americum-241 can be extracted from the ion chamber.*

Another experiment that can be conducted without radioactive sources is a charged balloon explained by (Austen & Brouwer, 1997).

First charging an inflated balloon with hair or skin. Then suspend the balloon using a tape for it to absorb radon particles and radon daughter nuclides that are present in the room. After 20 minutes, deflate the balloon and wrap it around the pin-diode. The count rate is significantly higher than the background radiation

## 3 Limitations

There were some problems that were encountered with earlier versions of the software which has already been rectified. Some known issues with the experiments include:

- length of pipe cannot be detected accurately in a noisy environment. Hence, forming larger groups may help in reducing the noise from other groups
- fluorescent light not available in the room
- Gas not in database identified wrongly
- alpha source cannot be detected by the pin-diode

## 4 Conclusions

The experiments were conducted in teacher professional development workshops and subsequently implemented in schools with students in Grades 11 and 12.

The methodology used by teachers varied across schools. Here are some ways that the observed lessons were implemented.
- Full experiment. Students were guided by worksheet to collect the data and analyze the results
- Experiment stations. Students move in groups from station to station doing different experiments. These experiments were implemented together with other experiments that shared similar concepts. These experiments were usually shorter than full experiment.
- Lecture demonstration. Students were taught the theory and teachers showed relevant experiments as a demonstration to illustrate the concept.

Both teachers and students' feedback were positive. They appreciated how the abstract concepts can be visualized using smartphones. They also appreciate that the web app doesn't require any installation and could be implemented quickly.

This paper shows how smartphones can be used to easily conduct experiments on standing sound waves, line spectrum, and radioactivity experiments. Details of the experimental setup and open source app is shared in the paper.

The apps are open source and we encourage everyone to further customize the app to suit your learning objectives.

## 5 Acknowledgments

## Appendix A: Spectrometer design

The 3-D printing file for the spectrometer can be accessed at gg.gg/spectrometer.

The design is modified by open source public lab to allow the smartphone camera to record sharp line spectrum.

To make production easier, the spectrometer was designed to be a single piece to minimize assembly. A 1000 lines diffraction grating sheet needs to be attached to the window of the spectrometer.

We added a groove so that a rubber band can be used to secure the spectrometer to the phone.

Other optimization includes varying slit width and angle of the spectrometer.

During our testing, some older phones have difficulty detecting the light sources from the gas lamp and we had to increase the slit width to allow more light to enter the phone camera. However, the resolution of the line is compromised.

---

[7] https://iwant2study.org/ospsg/index.php/890-innergy-award-2019